\documentclass[a4paper,11pt]{article}
\usepackage{pos}
\usepackage{verbatim}

\usepackage[]{xcolor}

\title{Strong isospin breaking in $S\!p\!(4)$ gauge theory}

\author*{Axel Maas}
\author{Fabian Zierler}

\affiliation{Institute of Physics, NAWI Graz, University of Graz,\\
  Universitätsplatz 5, 8010 Graz, Austria}

\emailAdd{axel.maas@uni-graz.at}
\emailAdd{fabian.zierler@uni-graz.at}

\abstract{A hidden strongly-interacting sector is a possible candidate for dark matter. One scenario consistent with astrophysical constraints is a QCD-like theory with gauge group Sp(4) and two fundamental non-degenerate fermions. We report exploratory investigations of the light spectrum and decay constants for this theory using lattice calculations.}

\FullConference{%
 The 38th International Symposium on Lattice Field Theory, LATTICE2021
  26th-30th July, 2021
  Zoom/Gather@Massachusetts Institute of Technology
}

\begin{document}
\maketitle

\section{Introduction}

The nature of dark matter (DM) is one the most challenging problems in current particle physics. Recently a class of strongly-interacting, QCD-like gauge theories with a number of massive fermions has gained some interest as dark matter candidates \cite{Hochberg:2014dra,Hochberg:2014kqa,Hochberg:2015vrg,Bernal:2019uqr}. In these theories dark matter is made up of the pseudo-Goldstones bosons associated with the breaking of the approximate chiral symmetry and dark matter can self-annihilate through $3\to 2$ processes in the early universe. In addition non-degenerate fermions are interesting because they can establish a mass hierarchy in the dark sector. As an example of such a sector we study here $Sp(4)$ gauge theory with two fundamental non-degenerate Dirac fermions in isolation. To the best of our knowledge this is the first time that strong isospin breaking has been studied in a non-$SU(N)$ gauge theory on the lattice. Here we present first, exploratory results for the masses and decay constant of the pseudoscalar and vector iso-nonsinglet mesons.

\section{Setup, global symmetries and observables}

The fundamental representation in any symplectic group is pseudo-real. This entails relations between particles and anti-particles and their Weyl components. Therefore this theory has a global $U(4)_I$ isospin symmetry in case of vanishing masses. It is broken by the axial anomaly down to $SU(4)_I$. Spontaneous chiral symmetry breaking and/or degenerate fermions break this symmetry further to $Sp(4)_I$. In case of non-degenerate masses the symmetry is broken down further to $SU(2)_u \times SU(2)_d$ \cite{Kosower:1984aw,Kulkarni:2021}. These relate the Weyl components of the two dark fermion flavours, the (dark) up and the (dark) down.

 As this theory confines the physical degrees of freedom are hadrons. Due to the relation between particles and anti-particles and the even number of colors all hadrons are bosons, and made from an even number of fermions. In particular, in addition to the usual quark-antiquark mesons additional diquark (quark-quark and antiquark-antiquark) bound-states appear. It can be shown that these states have the same correlation functions as their mesonic counterparts but have flipped parity \cite{Hands:2007uc,Lewis:2011zb}. We will therefore restrict ourselves to the study of quark-antiquark pseudoscalar and vector bound-states. Note, that the corresponding quark-quark states are, however, scalars and axialvectors.
 
 Most interesting for dark matter physics are the properties of the lightest states. These are the (pseudo)-Goldstone bosons as well as (likely) the nonsinglet vector mesons. In the mass degenerate case these quantities have already been studied in \cite{Bennett:2017kga,Bennett:2019jzz}.
 
 Because the global symmetry is further broken by the non-degenerate quark masses, these hadrons will no longer be degenerate: Similar to QCD, the flavoured and unflavoured mesons will obtain different masses and decay constants. In the present case the 5 Goldstones decompose into 2 fundamental representations of each of the SU(2) groups and one singlet. Likwise, the 10-plet of flavoured vector mesons \cite{Bennett:2019cxd} decompose into a 6-plet of flavoured particles and 4-plet of particles \cite{Kulkarni:2021}.

We will denote the unflavoured pseudoscalar by $\pi^0$ and the flavoured ones as $\pi^\pm$ in analogy to QCD. The other pair is made up of quarks, and remains degenerate to the $\pi$ \cite{Kulkarni:2021}. Similarly we consider  $\rho^\pm$ and the unflavoured ones $\rho^0$.

\section{Lattice setup}

We use a variant of the HiRep code \cite{DelDebbio:2008zf} adapted for $Sp(N)$ gauge theories \cite{Bennett:2017kga} to simulate $Sp(4)$ gauge theory using the standard Wilson gauge action. We include two fundamental Dirac-Wilson-fermions with non-degenerate masses using the RHMC algorithm. 
\begin{figure}
    \centering
    \includegraphics[width=.63\textwidth]{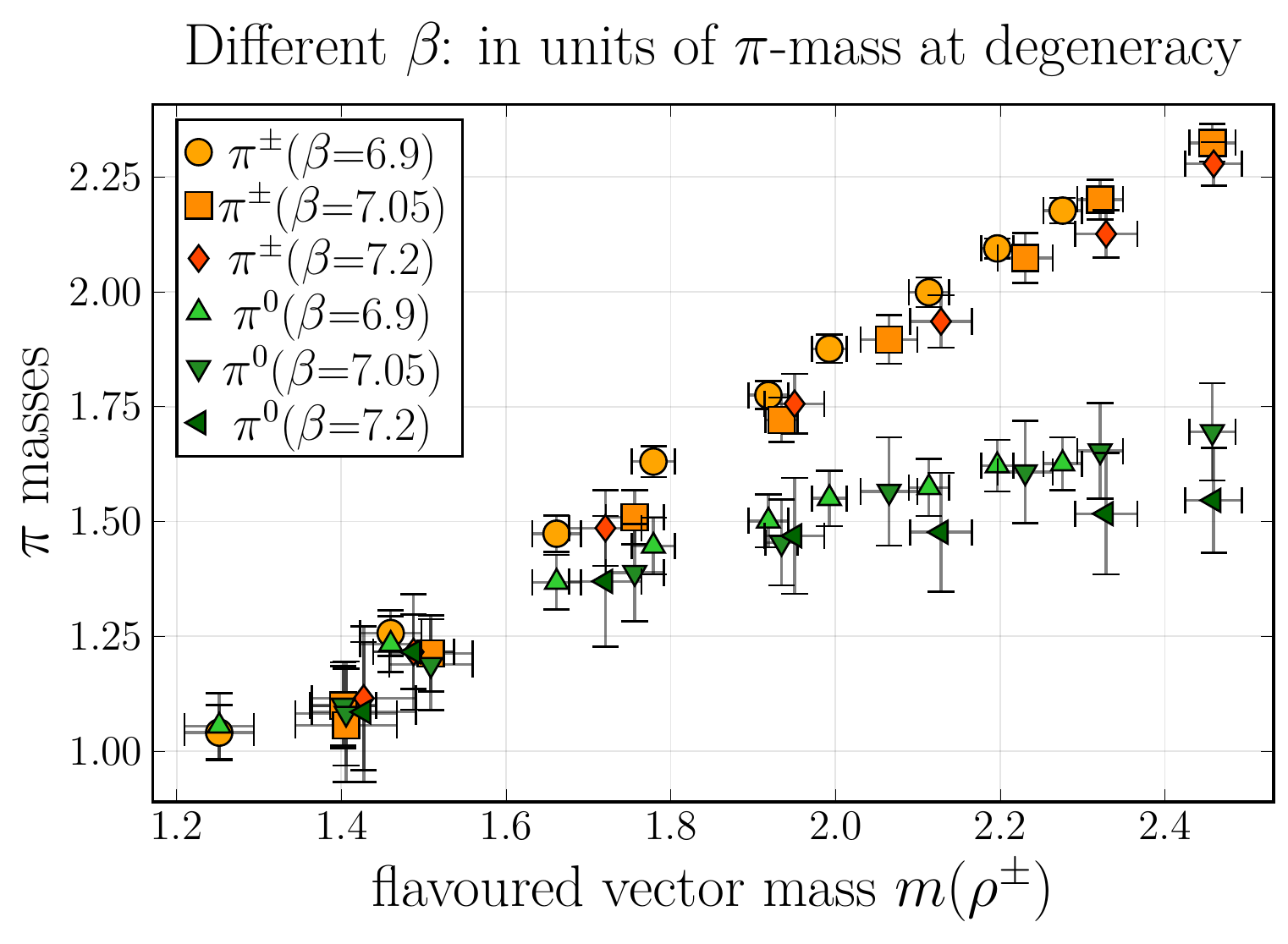}
    \includegraphics[width=.35\textwidth]{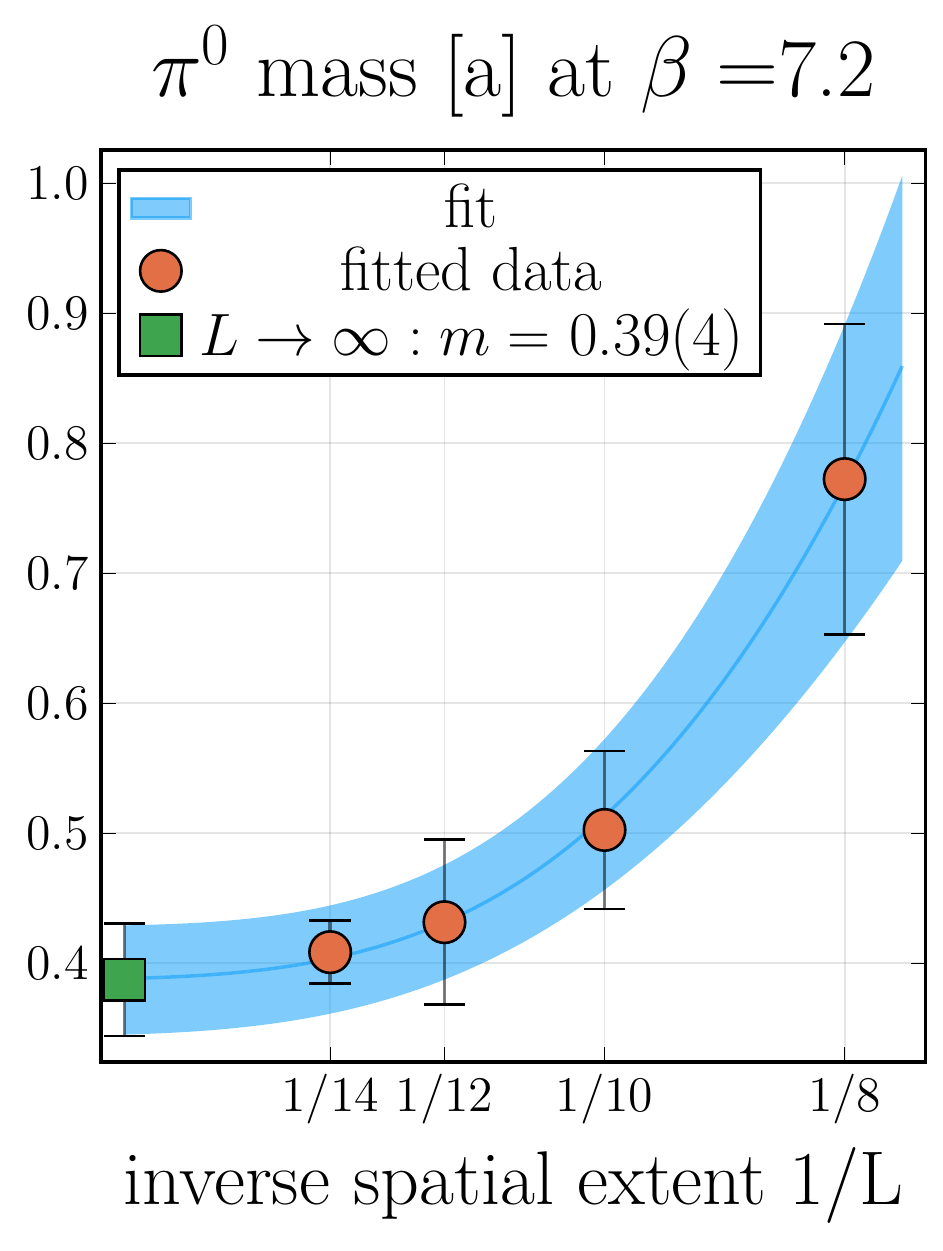}
    \caption{(left) Goldstone masses in units of the mass at degeneracy for all three different values of $\beta$. For all three different values the almost identical behaviour is observed. (right) Unflavoured Goldstone $\pi^0$ masses as a function of the spatial lattice extent $L$ for $\beta=7.2$ with the lightest set of quark masses considered. An extrapolation to infinite volume using the same fit function as in \cite{Bennett:2019jzz} is included. The value obtained on the largest lattice agrees within errors with the finite volume extrapolation. }
    \label{fig:systematics}
\end{figure}

We have chosen values for the inverse coupling $\beta > 6.8$ so that we avoid the bulk phase transition described in \cite{Bennett:2019jzz}. We have chosen three particular values of $\beta=6.9,7.05,7.2$. For all these values of the inverse coupling results for the mass degenerate case have already been obtained on larger lattices in \cite{Bennett:2019jzz}. We fix the lighter of the two quark masses by requiring an approximate ratio of $m_\rho / m_\pi \approx 1.25$ in the degenerate case. This is suggested by dark matter phenomenology \cite{Berlin:2018tvf}, which prefers stable vector particles and Goldstones of order or above the confinement scale. The other bare quark mass is then incrementally increased.

We considered lattices up to the size of $L^3 \times T = 14^3 \times 24$ such that the finite volume effects are under control. This is demonstrated in figure \ref{fig:systematics}. Additionally, we show in figure \ref{fig:systematics}
a comparison of the three different values of $\beta$. It can be seen that even in the case of the smallest lattice spacing the infinite volume limit and the result obtained on the largest lattice ($24 \times 14^3$) agree within errors. We report that in all ensembles studied the product of $m_{\pi} L$ was always $> 5.5$. However, at very large mass splittings lattice cutoff effects may become relevant, and will be studied in the future. Details will be provided elsewhere. 

Denoting the (dark) quark fields as $u$ and $d$ these hadrons are described by operators $O_\Gamma^f$, specifically
\begin{align}
	O_\Gamma^\pm(x,y) &= \bar u (x) \Gamma d(y) \\
	O_\Gamma^0(x,y) &= \bar u (x) \Gamma u(y) - \bar d (x) \Gamma d(y).
\end{align}

The decay constants are defined through the same parameterization of the matrix elements as in \cite{Bennett:2019jzz},
\begin{align}
	\langle 0 | O_{\gamma_5 \gamma_\mu}^f | PS \rangle &= f_\pi^f p_\mu & 
	\langle 0 | O_{\gamma_\mu}^f | V \rangle &= f_\rho^f m_\rho^f \epsilon_\mu
\end{align}
where $| PS \rangle$ and $| V \rangle$ denote the ground state of the relevant meson. Here $\epsilon_\mu$ is the polarisation vector which fulfills $\epsilon_\mu \epsilon^\mu = 1$ and $\epsilon_\mu p^\mu = 0$. The masses are extracted from the exponential fall-off of the zero-momentum correlator on a lattice of temporal extent $T$ and spatial extent $L$ at large times $t$
\begin{align}
	&C_O(\vec p, t_x - t_y) = \frac{1}{L^3} \sum_{\vec x, \vec y} e^{-i\vec p (\vec x - \vec y)} \langle O(\vec x, t_x) O^\dagger(\vec y, t_y)\rangle \\
	&C_{O_\Gamma}(0,t) \xrightarrow[]{t\to \infty}  \frac{|\langle 0 | O_\Gamma | GS  \rangle|^2}{2 m_\text{meson} L^3}  \left( e^{-m_\text{meson} t} + e^{-m_\text{meson} (T-t)} \right) 
\end{align} 
where $| GS \rangle$ is the ground state of the meson sourced by the interpolator $O_\Gamma$. 
In the rest frame the matrix element parameterizing the pseudoscalar decay constant becomes
\begin{align}
    \langle 0 | O_{\gamma_5 \gamma_0} | PS \rangle &= f_\pi m_\pi
\end{align}
and we can extract all desired information by fitting the correlators at large $t$ to 
\begin{align}
    C_{O_{\gamma_5 \gamma_0}}(0,t) &\xrightarrow[]{t\to \infty}  \frac{(f_\pi)^2 m_\pi }{2 L^3}  \left( e^{-m_\pi t} + e^{-m_\pi (T-t)} \right) \\
    C_{O_{\gamma_\mu}}(0,t) &\xrightarrow[]{t\to \infty}  \frac{(f_\rho)^2 m_\rho }{2 L^3}  \left( e^{-m_\rho t} + e^{-m_\rho (T-t)} \right).
\end{align}

\noindent For the unflavoured mesons disconnected contributions to the correlator appear that vanish exactly in the mass-degenerate limit. These are, however, suppressed by the relatively large quark masses and the small mass difference $(m_u\! - \!m_d)$ and are therefore for the moment neglected. 

\section{Results}

\begin{figure}
    \centering
    \includegraphics[width=.85\textwidth]{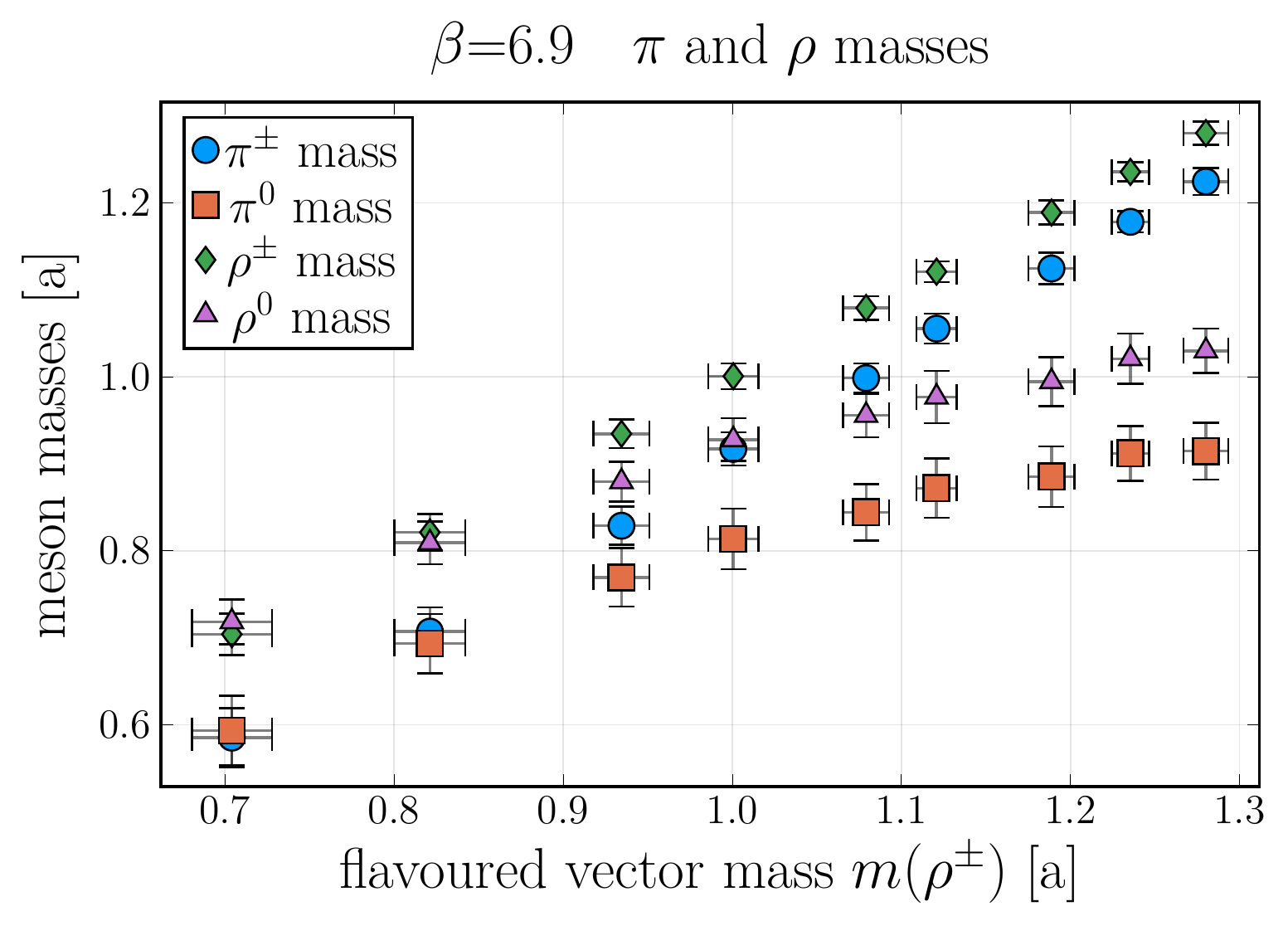}
    \caption{Masses of the flavoured and unflavoured Goldstone and iso-nonsinglet vector mesons masses in lattice units. At sufficiently large fermionic mass difference the unflavoured vectors get even lighter than the flavoured Goldstones.}
    \label{fig:masses}
\end{figure}

\begin{figure}
    \centering
    \includegraphics[width=.85\textwidth]{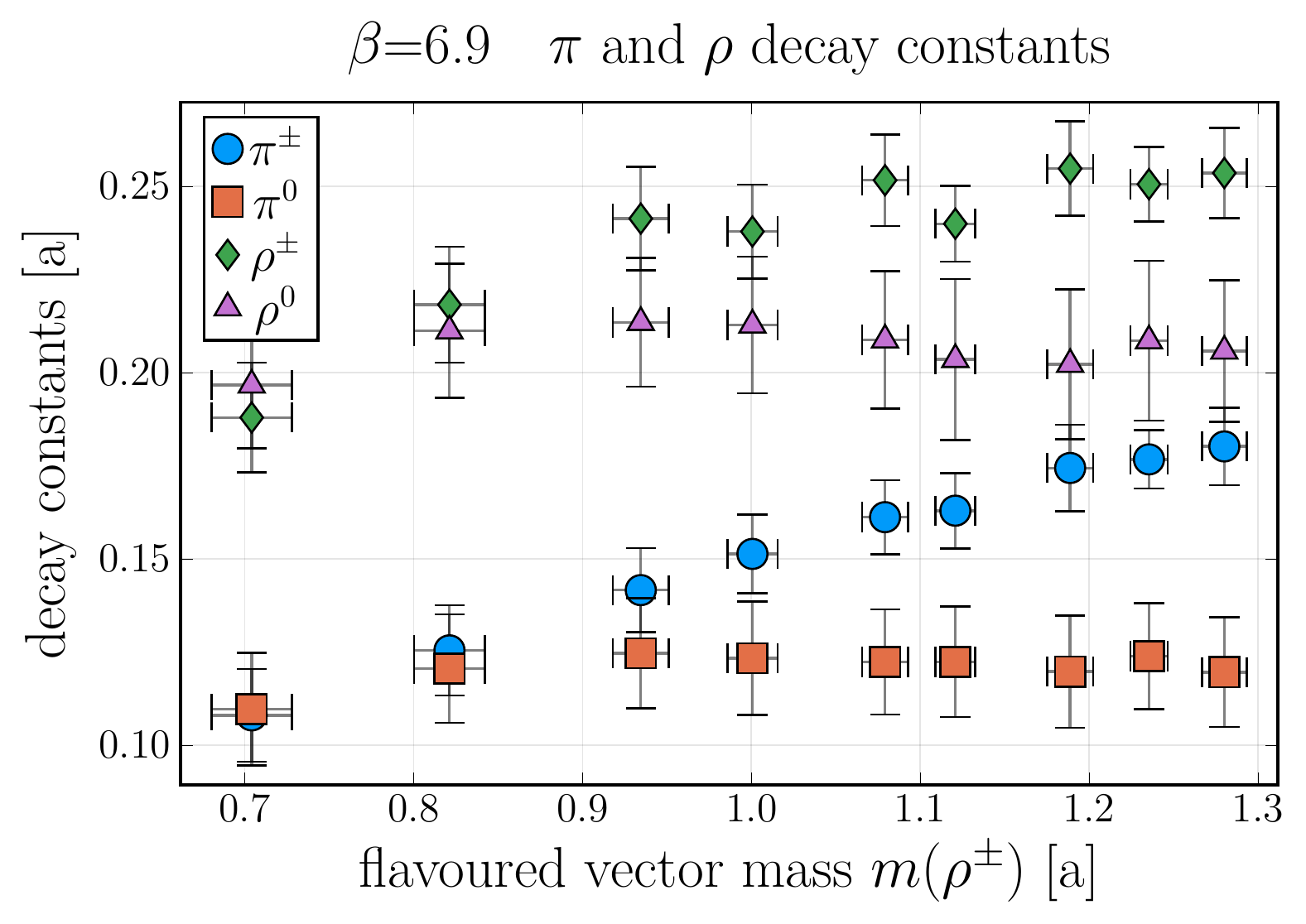}
    \caption{Decay constants of the flavoured and unflavoured Goldstone and iso-nonsinglet vector mesons in lattice units. For large mass differences of the fermions the decay constants show hints of non-monotonicity. }
    \label{fig:decayconstants}
\end{figure}

The results for the flavoured and unflavoured pseudo-Goldstone and vector meson masses and decay constants are depicted in figures \ref{fig:masses} and \ref{fig:decayconstants}. The results are given in lattice units. If desired, a scale can be introduced, e.\ g.\ by fixing the lattice spacing $a$.

We see a clear separation of the flavoured and unflavoured meson states and can conclude that they are no longer mass-degenerate. For both the pseudo-Goldstones and the vectors the unflavoured ones are lighter than the others. Furthermore, at sufficiently large fermion mass difference the unflavoured vectors get even lighter than the flavoured pseudo-Goldstones. Since one fermion mass is kept fixed while the other is incrementally increased all mesons get heavier the further we move away from degeneracy.

In the case of the form factors we again observe that the flavoured and unflavoured ones are no longer degenerate and the unflavoured form factors are smaller. In addition we see hints of non-monotonicity in both form factors.

An important observation is that we see clearly qualitatively different behaviors. Especially, the unflavoured ones do not show a (dominant) linear increase with the mass of one of the flavours. For the decay constants the effect is even different between flavoured  pseudoscalars and vectors. This strongly suggests that the composition of the dark hadrons changes substantially. This can have implications for decay patterns, which will be studied elsewhere \cite{Kulkarni:2021}. Note, however, that in our case even at the largest mass spitting the vector mesons remain stable.

\section{Summary}

We have studied $Sp(4)$ gauge theory with two non-degenerate massive fundamental Dirac fermions. Four of the five pseudo-Goldstone bosons remain mass-degenerate for $m_u \neq m_d$ while the unflavoured one aquires a lighter mass and is therefore the lightest state in this theory.

We see strong signs for non-linear behavior, which suggests that the flavour composition of the dark hadrons changes. This can have interesting consequences for phenomenology. Also, our results show that it is possible to extract the splitting effects sufficiently well that interpolation even to tiny mass splittings, as preferred in some dark matter scenarios \cite{Bernal:2017mqb}, would be accessible. A more detailed analysis of the phenomenological implications as well of lattice details is forthcoming \cite{Kulkarni:2021}.

\acknowledgments
FZ is supported by the the Austrian Science Fund research teams grant STRONG-DM (FG1). The computations have been performed on the HPC cluster of the University of Graz and on the Vienna Scientific Cluster (VSC4). We are grateful to the authors of \cite{Bennett:2017kga,Bennett:2019jzz} for the support with the HiRep code.


\bibliographystyle{JHEP}
\bibliography{bib.bib}


\end{document}